\documentclass[acmsmall,screen,nonacm]{acmart}
\renewcommand\footnotetextcopyrightpermission[1]{}

\setcopyright{none}

\usepackage{algorithmicx}
\usepackage{enumitem}
\usepackage{amsthm}
\theoremstyle{remark}
\newtheorem{remark}{Remark}

\usepackage{graphicx}%
\usepackage{multirow}%
\usepackage{amsmath}%
\allowdisplaybreaks
\usepackage{mathrsfs}%
\usepackage[title]{appendix}%
\usepackage{textcomp}%
\usepackage{manyfoot}%
\usepackage{booktabs}%
\usepackage{comment}
\usepackage{algorithm}%
\usepackage{algorithmicx}%
\usepackage{algpseudocode}%
\usepackage{microtype}
\usepackage{caption}
\usepackage{subcaption}
\usepackage{csquotes}
\usepackage{listings}
\usepackage{isabelle/isabelle, isabelle/isabellesym}
\usepackage{kotex}
\usepackage{xcolor}
\usepackage[most]{tcolorbox}
\usepackage{wrapfig}
\definecolor{emerald}{RGB}{26,121,42}
\definecolor{topaz}{RGB}{236,185,57}
\definecolor{sapphire}{RGB}{14,26,164}
\usepackage{tabularx}
\newcolumntype{Y}{>{\centering\arraybackslash}X}
\definecolor{isa-bg}{RGB}{248,248,246}
\definecolor{isa-key}{RGB}{34,139,34}
\definecolor{isa-cmt}{RGB}{90,100,110}

\lstdefinelanguage{Isabelle}{
  morekeywords={
    theory,imports,begin,end,lemma,theorem,definition,
    unfolding,by,proof,qed,assumes,shows,if,then,else, apply, done
  },
  sensitive=true,
  morecomment=[n]{(*}{*)},
  morestring=[b]",
  alsoletter={\\<>},
  literate=
    {\\<lbrace>}{{$\lbrace$}}1
    {\\<rbrace>}{{$\rbrace$}}1
    {\\<lambda>}{{$\lambda$}}1
    {\\<noteq>}{{$\neq$}}1
    {\\<longrightarrow>}{{$\longrightarrow$}}1
    {\\<or>}{{$\vee$}}1
}

\lstdefinestyle{isabelle}{
  language=Isabelle,
  basicstyle=\ttfamily\small,
  keywordstyle=\bfseries\color{isa-key},
  backgroundcolor=\color{isa-bg},
  columns=fullflexible,
  keepspaces=true,
  breaklines=true,
  showstringspaces=false,
  frame=single
}

\AtBeginDocument{%
  }

\presetkeys{todonotes}{inline, color=topaz!40, bordercolor=topaz}{}

\newcommand{\sysname}{PROMISE} 
\newcommand{\sel}{seL4}
\newcommand{\selene}{Selene}
\newcommand{\rango}{Rango}

\usepackage{pifont}
\newcommand{\cmark}{\ding{51}}
\newcommand{\xmark}{\ding{53}} 
\usepackage[normalem]{ulem}

\begin{document}

\title{\sysname: Proof Automation as Structural Imitation of Human Reasoning}

\author{Youngjoo Ahn}
\affiliation{%
  \institution{Yonsei University}
  \city{Seoul}
  \country{South Korea}
}
\email{lordahn@yonsei.ac.kr}

\author{Sangyeop Yeo}
\affiliation{%
  \institution{Yonsei University}
  \city{Seoul}
  \country{South Korea}
}
\email{sangyeop@yonsei.ac.kr}

\author{Gijung Im}
\affiliation{%
  \institution{Yonsei University}
  \city{Seoul}
  \country{South Korea}
}
\email{kijeonglim@yonsei.ac.kr}

\author{Jongmin Lee}
\affiliation{%
  \institution{Yonsei University}
  \city{Seoul}
  \country{South Korea}
}
\email{jongmin.lee@yonsei.ac.kr}

\author{Jinyoung Yeo}
\affiliation{%
  \institution{Yonsei University}
  \city{Seoul}
  \country{South Korea}
}
\email{jinyeo@yonsei.ac.kr}

\author{Jieung Kim}
\affiliation{%
  \institution{Yonsei University}
  \city{Seoul}
  \country{South Korea}
}
\email{jieungkim@yonsei.ac.kr}

\renewcommand{\shortauthors}{Ahn et al.}

\newcommand{\jmlee}[1]{\textcolor{red}{#1}}

\begin{abstract}
Automated proof generation for formal software verification remains a largely unresolved challenge despite recent advances in large language models (LLMs). Generative AI has achieved remarkable success across domains such as natural language processing, computer vision, and software engineering tasks including code generation and summarization. However, several labor-intensive tasks that require substantial human expertise have not yet benefited significantly from these models. Among them, formal verification is one of the well-known domains in which LLMs are still struggling with. Formal verification has a huge benefit that it can provide machine-checked guarantees of functional correctness for safety-critical systems. Despite this benefit, however, its adoption remains limited due to scalability issues. In practice, the process relies on interactive theorem proving (ITP), where engineers must manually construct detailed proofs under strict logical constraints, requiring significant expertise and time.  For example, verifying the seL4 microkernel, which is the first fully verified operating system kernel, required decades of cumulative person-years of effort.  Consequently, recent advances in LLMs have renewed interest in reducing this human burden through automated proof synthesis. Nevertheless, existing LLM-assisted approaches remain fundamentally limited. Most treat proof generation as either a single-shot generation task or a shallow retrieval-augmented process, relying on repeated sampling to accumulate successful attempts. Such strategies work to some extent when the proof context is relatively small. However, they rapidly struggle to scale to large, highly interdependent verification artifacts, such as operating systems, microkernels, or verified compilers, where proofs exhibit deep structural dependencies across files, modules, and abstraction layers.

To address this challenge, we present \sysname\ (\textbf{PRO}of \textbf{MI}ning via \textbf{S}tructural \textbf{E}mbeddings), a structure-aware proof mining framework that reframes proof generation as a stateful and iterative search process guided by \textit{structurally similar proofs}. Rather than retrieving premises or proofs based on surface-level textual similarity, \sysname\ mines structural similarity over proof states and tactic-transition traces, capturing how goals, assumptions, and contexts evolve during successful derivations. By aligning proof-state transitions, \sysname\ retrieves structurally compatible proof fragments and incrementally adapts them to new targets, transforming repeated LLM trials into guided, state-aware proof search. We evaluate \sysname\ on the seL4 microkernel verification benchmark using multiple LLM backends, including GPT-3.5-Turbo and Qwen2.5-Coder-7B-Instruct, and compare it with recent LLM-based proof automation systems such as \selene~\cite{zhang-etal-2024-selene} and \rango~\cite{10.1109/ICSE55347.2025.00161}. \sysname\ consistently outperforms prior approaches across the majority of settings, with improvements reaching up to +26 points (186\% relative gain), while remaining competitive with the strongest baseline in the sole exception (GPT-4.1/P2).  Moreover, \sysname\ maintains strong performance across models of varying capability when compared with other methods. These results demonstrate that structure-aware mining of proof-state transition traces is an important step toward making LLM-assisted theorem proving more robust and scalable for real-world system software verification.
\end{abstract}


\maketitle


\section{Introduction}\label{sec:intro}
Formal verification provides the strongest guarantees of software correctness. Unlike testing, which can only reveal the presence of bugs, formal verification enables mathematically proving that a system satisfies its specification. Due to these benefits, formal verification has been actively investigated over the past several decades~\cite{10.1561/2500000045}, both through improvements in the underlying verification theories and through the application of these theories to increasingly large and complex systems. As a result, several landmark verification projects have demonstrated that full functional correctness of complex system software is achievable. Notable examples include the seL4 microkernel~\cite{Klein_EHACDEEKNSTW_09}, the FSCQ verified file system~\cite{chen2015fscq}, and the CertiKOS certified operating system~\cite{10.1145/2103799.2103803}. These efforts provide clear evidence that large-scale system software can be verified end-to-end with strong mathematical guarantees.

However, these successes come at a substantial cost. Large verification projects often require years of research and engineering effort by expert teams, and the resulting proof artifacts frequently exceed the size of the implementation itself. For example, the CertiKOS project contains nearly 200,000 lines of proofs, whereas the verified kernel code consists of fewer than 7,000 lines~\cite{10.1145/2103799.2103803}. Similar patterns have been observed in other large-scale verification projects. Their proof developments involve tens of thousands of lemmas and complex dependencies across modules and abstraction layers. These observations highlight a fundamental scalability challenge of formal verification. While verification can guarantee correctness, proof construction remains extremely labor-intensive.

Beyond their size, large verification developments show structural characteristics that make proof construction particularly challenging. Proof artifacts often form deep dependency graphs across files and modules, where the proof of a high-level theorem depends on long chains of auxiliary lemmas and library results. Moreover, reasoning patterns frequently reappear across different components of the system, even when the concrete lemmas and contexts differ. These recurring reasoning motifs include sequences of rewriting, simplification, invariant preservation, and rule applications that collectively form characteristic proof evolution patterns. Managing such large and interconnected proof structures requires not only logical correctness but also significant engineering effort to organize and reuse reasoning patterns across proofs. Consequently, reducing the cost of proof construction has become a central challenge in formal verification research.

Consequently, LLM-based automated proof generation has recently emerged as one of the most promising directions for reducing the cost of proof construction. While earlier proof automation techniques relied primarily on program logics and specialized tactics, LLMs introduce a fundamentally different paradigm based on proof synthesis through large-scale learned reasoning. When integrated with trusted proof assistants, LLMs can synthesize tactics~\cite{Cobblestone}, generate intermediate proof steps~\cite{10.1109/ICSE55347.2025.00161}, and sometimes construct entire proofs~\cite{Baldur, AutoVerus}. These approaches have shown promising results in mathematical theorem proving and small-scale program verification tasks, where reasoning contexts remain relatively localized. Building on these successes, recent studies have begun exploring whether LLMs can assist in verifying large system software. Benchmarks such as Selene investigate the feasibility of applying LLM-based proof generation to the seL4 ecosystem~\cite{zhang-etal-2024-selene}, while experiments on the FSCQ file system demonstrate that LLMs can automatically generate proofs for a nontrivial subset of system-level lemmas~\cite{10.1145/3713082.3730382}. However, despite these encouraging results, current LLM-assisted proof generation methods remain far from scaling to large verification developments. Empirical evidence shows that their effectiveness degrades sharply on large, highly interdependent proof artifacts. On system-scale benchmarks such as \sel\ or FSCQ, reported proof coverage for mid-level lemmas often remains below 30\%, even when using state-of-the-art models~\cite{zhang-etal-2024-selene,10.1145/3713082.3730382}. These observations suggest that simply improving model capability may not be sufficient to overcome the scalability limitations of current approaches.

\begin{figure*}
    \centering
    \includegraphics[width=1\linewidth]{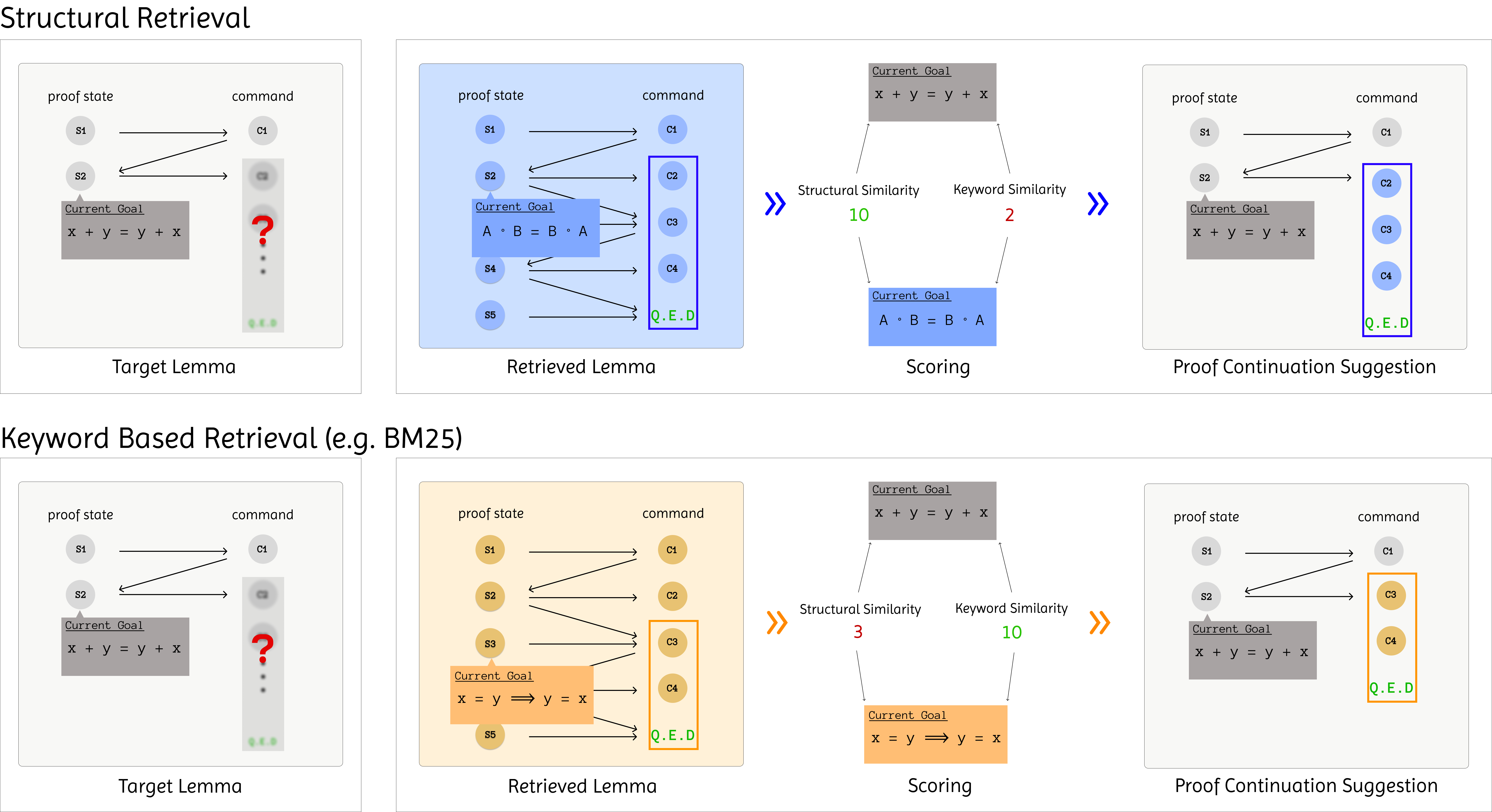}
    \caption{Structural retrieval over proof-state transition traces. Unlike keyword-based retrieval that operates on lemmas or proofs, \sysname\ retrieves structurally compatible proof fragments based on proof-state evolution patterns.}
    \label{fig:overview}
\end{figure*}
\begin{table}[H]
\centering
\small
\begin{tabular}{lcccc}
\toprule
\multirow{2}{*}{System} & Retrieval  & Reasoning   & Structural  & Proof-State \\
                        & Unit       & Granularity & Reasoning   & Aware \\
\midrule
LeanDojo & Premises & Lemma-level & \xmark{} & \cmark{} \\
\rango & Proofs + Lemmas & Lemma / tactic & \xmark{}  & \cmark{} \\
Selene & Artifacts & Proof-level & \xmark{} & \xmark{}  \\
\textbf{\sysname\ (Ours)} & \textbf{Proof-state transitions} & \textbf{Transition patterns} & \cmark{} & \cmark{} \\
\bottomrule
\end{tabular}
\caption{Comparison of retrieval strategies in LLM-assisted proof generation. 
Existing approaches primarily retrieve static artifacts such as premises or proofs, 
whereas \sysname\ retrieves proof-state transition traces to capture structural reasoning patterns.}
\label{tab:intro-comparison}
\end{table}
A key reason for these limitations lies in how proof automation is formulated in existing approaches. Most current methods treat proof generation either as a search problem over tactic sequences or as a text-generation task augmented with retrieval of relevant artifacts such as premises, lemmas, or prior proofs. Under this formulation, proofs are treated largely as independent artifacts, and knowledge reuse is limited to static textual objects retrieved on a per-proof basis. While such keyword-based retrieval can provide useful contextual hints, it often fails to capture the structural dynamics of proof construction in large verification developments. In practice, both automated systems and human proof engineers construct proofs through sequences of incremental transformations of proof states, where each step modifies the goal structure, introduces new assumptions, or reduces the remaining subgoals. These transformations form characteristic proof-state transition patterns that recur across different proofs. However, when retrieval is performed solely at the level of entire lemmas or proof scripts, these finer-grained structural relationships are often lost.


Our intuition originates from how human proof engineers actually construct proofs. Rather than relying on surface-level syntactic similarity between the current goal and its context, they reason about how the proof structure evolves over time. They also select lemmas based on whether their semantic roles align with the current proof state, rather than merely matching tokens or textual similarity. From a data perspective, large verification projects naturally expose this behavior through rich traces of proof-state transitions, capturing how goals, assumptions, and contexts evolve during successful derivations. These traces encode both structural information in proof progression and implicit signals about lemma applicability. This observation suggests that effective proof reuse requires jointly modeling (1) structural proof evolution and (2) context-sensitive, semantically aligned lemma usage. 


Building on this observation, we develop \sysname, a structure-aware proof generation framework that explicitly captures and leverages both aspects. First, \sysname\ models proofs as sequences of proof-state transitions and performs structural retrieval to identify previously observed reasoning paths that are analogous to the current proof state. This enables the system to reuse how proofs progress, rather than merely what artifacts they contain. Second, to support semantically appropriate lemma selection, \sysname\ integrates context-aware lemma retrieval by extracting the set of available lemmas from the proof environment (e.g., via Isabelle PIDE) and selecting those whose functional roles align with the current goal state. These two components are tightly coupled during generation: structural signals guide the reasoning trajectory, while semantically relevant lemmas ground each step in valid logical transformations. Importantly, \sysname\ is model-agnostic and operates with off-the-shelf language models without additional training, enabling consistent application across different LLM backends while preserving a unified, state-aware proof-generation strategy.

To evaluate the effectiveness of this approach, we conduct an empirical study comparing \sysname\ with existing LLM-assisted proof generation methods. Our evaluation focuses on two key aspects: absolute proof-generation performance and sensitivity to variations in the underlying language model. Under the same query budget, \sysname\ consistently achieves substantially higher proof success rates than prior approaches. Moreover, the results indicate that the proposed framework is significantly less sensitive to the choice of LLM backend. In particular, the performance difference between Qwen2.5-Coder-7B-Instruct~\cite{Qwen2.5-Coder-7B-Instruct} and GPT-3.5-Turbo remains minimal within our framework, whereas existing approaches typically exhibit large performance gaps across models. These results suggest that leveraging structural information from proof-state transition traces provides a more robust and scalable foundation for LLM-assisted theorem proving in large verification developments.

Our contributions are summarized as follows:

\begin{itemize}

\item \textbf{Structural view of proof generation.}
We introduce a new perspective that frames automated proof generation as a process-level reasoning problem over proof-state transitions, rather than as keyword-based retrieval or tactic prediction. This view identifies structural proof evolution as a first-class signal for reasoning in large-scale verification.

\item \textbf{Structure-aware proof generation framework.}
Building on this perspective, we develop \sysname\ (\textbf{PRO}of \textbf{MI}ning via \textbf{S}tructural \textbf{E}mbeddings), a model-independent framework that operationalizes structural reasoning by retrieving and leveraging proof-state transition patterns to guide state-aware proof search.

\item \textbf{Comprehensive empirical evaluation.}
We conduct an extensive evaluation of \sysname\ on system-scale verification benchmarks derived from seL4, comparing against prior approaches (Selene) under the same query budget. Our results show that \sysname\ consistently achieves higher proof success rates while remaining robust across different LLM backends, including Qwen2.5-Coder-7B-Instruct and GPT-3.5-Turbo.

\end{itemize}

The remainder of this paper is organized as follows. Section~\ref{sec:background} presents the background and motivation for LLM-assisted proof generation in large verification developments. Section~\ref{sec:methodology} describes the design and methodology of \sysname. Section~\ref{sec:evaluation-setup} introduces the evaluation setup, including benchmarks and evaluation metrics. Section~\ref{sec:evaluation} presents our experimental results and discussion. Section~\ref{sec:related-works} reviews related work, and Section~\ref{sec:conclusion} concludes the paper. All auto generated proofs associated with this work are publicly available at \url{https://zenodo.org/records/19049360}.
\section{Background and Motivation}\label{sec:background}

\subsection{Interactive Theorem Proving and Proof-State Transitions}

Interactive theorem proving (ITP) is a widely used paradigm for constructing mechanically verified proofs in systems such as Isabelle/HOL~\cite{IsabelleHOL}, Rocq/Coq~\cite{CoqArt}, and Lean 4~\cite{Lean4}. In these environments, proofs are developed incrementally through interaction with a proof assistant, which maintains a formal representation of the current proof state. A \emph{proof state} typically consists of two components: a set of goals and a proof context. The set of goals represents logical obligations that remain to be proven, while the context contains local assumptions, definitions, and previously introduced variables. Proof construction proceeds by applying \emph{tactics}, which transform the current proof state into one or more new proof states. Each tactic application may simplify the current goal, introduce new subgoals, or discharge existing ones.

From this perspective, proof development can be viewed as a sequence of proof-state transitions induced by tactic applications:
\[
s_0 \xrightarrow{t_1} s_1 \xrightarrow{t_2} s_2 \rightarrow \dots \rightarrow s_n
\]
where $s_0$ denotes the initial proof state derived from the theorem statement and $s_n$ denotes the terminal state in which all goals have been discharged. Instead of directly constructing proof terms, proof engineers typically write \emph{proof scripts}, which are sequences of tactic applications that guide the proof assistant in generating a complete formal proof object.

This state-transition view provides a useful abstraction for understanding how proofs evolve during reasoning. In practice, proofs rarely consist of a single reasoning step. Rather, they are constructed through sequences of incremental transformations of proof states, where each step modifies the goal structure, introduces new assumptions, or reduces the remaining subgoals. Importantly, many proofs share similar proof-state transition patterns. For example, reasoning sequences such as rewriting followed by simplification, case analysis followed by rule application, or introduction of assumptions followed by automated reasoning frequently recur across different proofs. These recurring transition patterns capture reusable reasoning strategies rather than specific proof artifacts. Consequently, understanding proof construction at the level of proof-state evolution provides a useful perspective for analyzing and supporting large-scale formal verification.

\subsection{Large-Scale Verification Proofs}

\begin{wrapfigure}{r}{0.50\textwidth}
    \centering
    \includegraphics[width=0.45\textwidth]{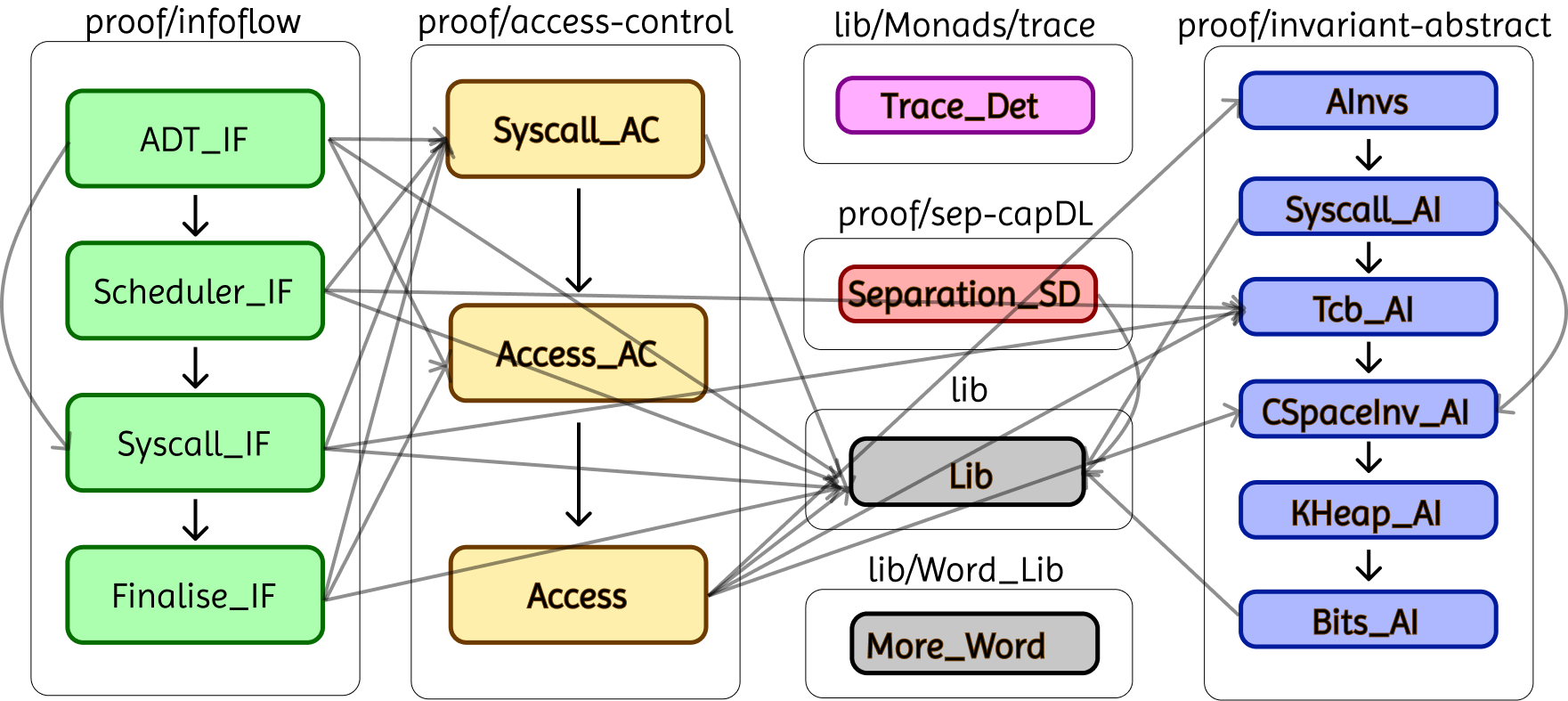}
    \caption{Benchmark Dependency Graph}
    \label{fig:sel4-dependency}
\end{wrapfigure}

While the proof-state transition model provides a general abstraction for interactive theorem proving, proofs in large-scale software verification projects show several distinctive characteristics that significantly complicate automated proof construction.

First, large verification developments contain extensive dependency graphs among definitions, lemmas, and theorems. A typical proof obligation rarely stands alone; instead, it relies on long chains of supporting results distributed across multiple modules and files. In industrial-scale verification projects such as seL4, FSCQ, and CertiKOS, proof developments consist of thousands of lemmas connected through deep inter-file dependencies. For instance, the seL4 verification project contains more than 100,000 lemmas and theorems, while the CertiKOS verification codebase—together with the CompCert libraries it relies on—contains hundreds of thousands of lines of specifications and proofs. Figure~\ref{fig:sel4-dependency} illustrates a portion of the dependency graph extracted from the seL4 verification project, which also serves as the primary target of our methodology. The figure highlights how individual lemmas are connected through long chains of dependencies across modules. Such complex dependency structures make automated proof construction particularly challenging, as identifying relevant prerequisites and supporting lemmas becomes nontrivial. 

Second, many verification proofs in their code base contain a lot of recurring reasoning. Typical examples include rewriting sequences, invariant preservation arguments, and structural reasoning about program states or data structures. These recurring reasoning motifs appear repeatedly across different components of the verification project. However, proof reuse rarely occurs at the level of entire proofs. Instead, proof engineers frequently reuse \emph{small proof fragments}, which correspond to short sequences of tactics implementing a particular reasoning step. For example, fragments such as \texttt{rewrite $\rightarrow$ simpl $\rightarrow$ apply lemma} or \texttt{intros $\rightarrow$ destruct $\rightarrow$ auto} commonly appear as parts of larger proofs. These fragments capture localized reasoning strategies that can be adapted across multiple proofs.

Finally, the applicability of such proof fragments depends heavily on the \emph{local proof context}. Tactic selection is influenced by several contextual factors, including the current goal, available assumptions, and the set of lemmas currently in scope. Consequently, two proof states that appear syntactically similar may require different reasoning steps depending on subtle contextual differences. This strong context sensitivity makes automated proof generation particularly challenging in large-scale verification environments, where both dependency structures and local reasoning contexts must be considered simultaneously.

Taken together, these characteristics indicate that scaling proof automation to large verification developments requires additional structural guidance beyond the raw generative capabilities of large language models.

\subsection{LLM-Assisted Proof Generation and Retrieval Limitations}

Recent advances in large language models (LLMs) have renewed interest in automated theorem proving. A growing body of work investigates how LLMs can assist proof development by generating tactics, constructing proofs, or providing proof guidance within interactive theorem proving environments. Existing approaches can be broadly categorized into three paradigms: tactic prediction, whole-proof generation, and retrieval-augmented proving.

\paragraph{\textbf{Tactic prediction.}}

The most common formulation treats proof generation as a next-tactic prediction problem. Given the current proof state $s$, the model predicts the next tactic $t$, yielding a mapping $s \rightarrow t$. This formulation enables incremental proof construction under proof assistant feedback and allows the model to interact with the proof assistant during the proof search process.

\paragraph{\textbf{Whole-proof generation.}}

Another line of work attempts to generate an entire proof directly from the theorem statement, which can be viewed as a mapping $\text{problem} \rightarrow \text{proof}$. While this approach has shown promising results for small mathematical benchmarks, it typically struggles with complex verification tasks due to long proofs, deep lemma dependencies, and strong context sensitivity.

\paragraph{\textbf{Retrieval-augmented proving.}}

More recent systems incorporate retrieval mechanisms to provide relevant context during proof generation. These systems retrieve artifacts such as lemmas, prior proofs, or proof fragments and provide them as contextual guidance to the language model during generation. By incorporating project-specific knowledge, retrieval-augmented approaches often improve proof generation performance in realistic verification environments.

Despite this progress, most existing retrieval mechanisms operate primarily at the level of static artifacts such as lemmas or proof fragments. However, these artifacts do not faithfully capture the true unit of reasoning reuse in proof development. In practice, reusable reasoning often manifests through proof fragments that correspond to short sequences of tactic applications and recurring proof-state transition patterns. Consequently, keyword-based retrieval frequently fails to capture the structural regularities that arise across proofs.

This limitation becomes particularly pronounced in large-scale verification environments, where many proofs share similar structural patterns but differ syntactically due to contextual variations. As a result, artifact-based retrieval often returns information that is either too coarse-grained or poorly aligned with the evolving proof state.

Table~\ref{tab:retrieval-comparison} summarizes representative retrieval strategies used in LLM-assisted proof generation systems. While existing approaches incorporate retrieval signals to improve generation, they primarily operate at the level of individual artifacts and lack mechanisms for capturing structural relationships across proof-state transitions.

\begin{table*}[t]
\scriptsize
\centering
\begin{tabular}{l l l}
\toprule
Category
& Retrieval Unit
\& Relevance Signal
& Key Limitation \\
\midrule


Neural / Embedding Retrieval~\cite{DBLP:conf/icml/BlaauwbroekORMP24, DBLP:conf/iclr/MikulaTAPJZSKMW24}
& Lemmas / proof states, Embedding or graph similarity 
& Lacks structural history \\

RAG-style Retrieval for LLMs~\cite{NEURIPS2023_44414694,zhang-etal-2024-selene}
& Lemmas / proof fragments, Goal–context similarity 
& Limited structural guidance \\

Retrieval + Proof Search~\cite{10.1109/ICSE55347.2025.00161}
& Lemmas + tactics, Relevance + search score 
& No cross-proof reuse \\

\uline{This work (\sysname)} 
& \uline{Proof-state transition traces, Structural + dependency signals} 
& -- \\

\bottomrule
\end{tabular}
\caption{Comparison of retrieval paradigms used in automated proof generation systems. Existing approaches primarily retrieve static artifacts such as lemmas or proof fragments, whereas \sysname\ retrieves proof-state transition traces to capture structural reasoning patterns across proofs.}
\label{tab:retrieval-comparison}
\end{table*}

\subsection{Motivating Example}

To describe the limitations of keyword-based retrieval in LLM-assisted proof generation, we present a motivating example from the seL4 verification project. This example compares the retrieval behavior of \sysname\ with that of an artifact-based retrieval approach. Listing~\ref{lst:akernel-invs-target} shows the target lemma \texttt{akernel\_invs} from the \texttt{AInvs} session of the seL4 verification project. \sysname\ successfully generates a correct proof for this lemma. In contrast, an artifact-based retrieval approach following the \textsc{Rango} framework fails to produce a valid proof within its rollout budget. The key difference lies in how the two systems retrieve supporting proofs during generation.

\sysname\ retrieves the structurally similar lemma shown in Listing~\ref{lst:akernel-invs-det-ext}. The Hoare-triple structure of this lemma closely matches that of the target lemma, and its proof follows the same reasoning skeleton: unfolding \texttt{call\_kernel} and discharging the goal using \texttt{activate\_invs} and \texttt{active\_from\_running}. Because the retrieved example aligns with the current proof-state transition, it provides an effective template for the language model to adapt. In contrast, the approach based on Rango retrieves the lemma shown in Listing~\ref{lst:call-kernel-valid-pdpt}. Although this lemma shares lexical similarity with the target, its proof structure differs substantially. The proof begins with a case split over \texttt{e} and involves a considerably more complex tactic sequence combining \texttt{wp}, \texttt{simp}, and \texttt{wpc}. In practice, this causes the language model to repeatedly generate malformed variants of this heavier reasoning pattern, ultimately exhausting the rollout budget without producing a valid proof.

This example highlights a key limitation of keyword-based retrieval. Textual or keyword similarity may identify related lemmas, but it does not necessarily capture the structural reasoning patterns required to complete the proof. In contrast, retrieving structurally aligned proof-state transitions provides more reliable guidance for proof generation, motivating the structure-aware retrieval strategy employed by \sysname.

\begin{lstlisting}[style=isabelle,
  caption={Target lemma from the AInvs session},
  captionpos=b,
  label={lst:akernel-invs-target}]
lemma akernel_invs:
  "\<lbrace>invs and (\<lambda>s. e \<noteq> Interrupt \<longrightarrow> ct_running s)\<rbrace>
  (call_kernel e) :: (unit,unit) s_monad
  \<lbrace>\<lambda>rv. invs and (\<lambda>s. ct_running s \<or> ct_idle s)\<rbrace>"
  unfolding call_kernel_def
  by (wpsimp wp: activate_invs simp: active_from_running)
\end{lstlisting}

\begin{lstlisting}[style=isabelle,
  caption={Top structural few-shot example retrieved by \sysname},
  captionpos=b,
  label={lst:akernel-invs-det-ext}]
lemma akernel_invs_det_ext:
  "\<lbrace>invs and (\<lambda>s. e \<noteq> Interrupt \<longrightarrow> ct_running s)\<rbrace>
  (call_kernel e) :: (unit,det_ext) s_monad
  \<lbrace>\<lambda>rv. invs and (\<lambda>s. ct_running s \<or> ct_idle s)\<rbrace>"
  unfolding call_kernel_def
  by (wpsimp wp: activate_invs simp: active_from_running)
\end{lstlisting}

\begin{lstlisting}[style=isabelle,
  caption={Top proof retrieved by \textsc{rango}},
  captionpos=b,
  label={lst:call-kernel-valid-pdpt}]
lemma call_kernel_valid_pdpt[wp]:
  "\<lbrace>invs and (\<lambda>s. e \<noteq> Interrupt \<longrightarrow> ct_running s) and valid_pdpt_objs\<rbrace>
   call_kernel e
   \<lbrace>\<lambda>_. valid_pdpt_objs\<rbrace>"
  apply (cases e, simp_all add: call_kernel_def)
      apply (rule hoare_pre)
       apply (wp | simp add: if_apply_def2 | wpc
                 | rule conjI | clarsimp simp: ct_in_state_def
                 | erule pred_tcb_weakenE
                 | wp (once) hoare_drop_imps)+
  done
\end{lstlisting}

\section{Methodology}\label{sec:methodology}
\subsection{PROMISE}
\begin{figure}[h]
    \centering
    \includegraphics[width=1\linewidth]{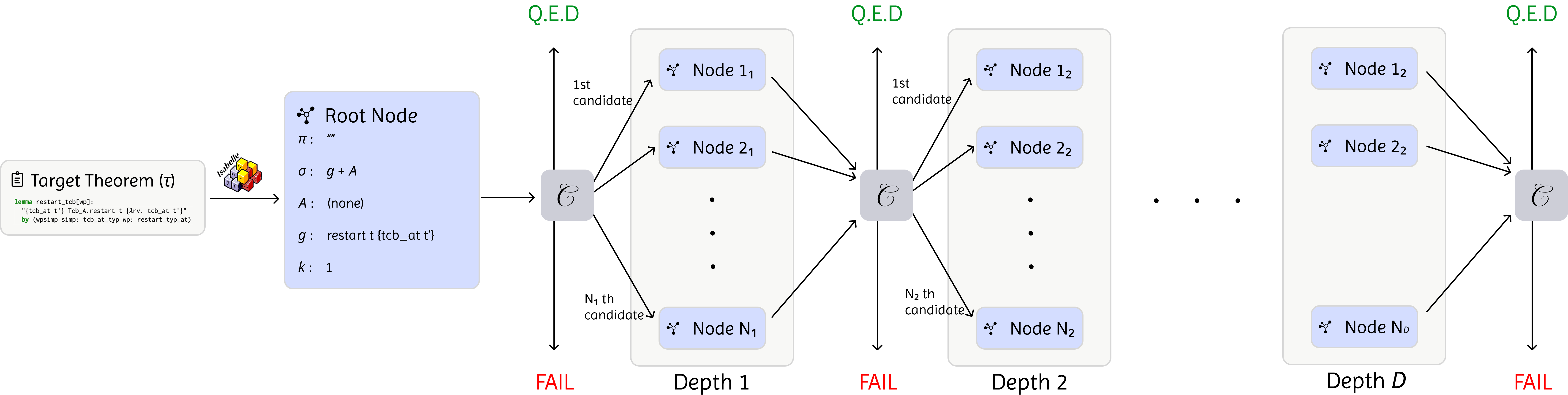}
    \caption{PROMISE Main Pipeline}
    \label{fig:main-pipeline}
\end{figure}
\begin{figure} [h]
    \centering
    \includegraphics[width=1\linewidth]{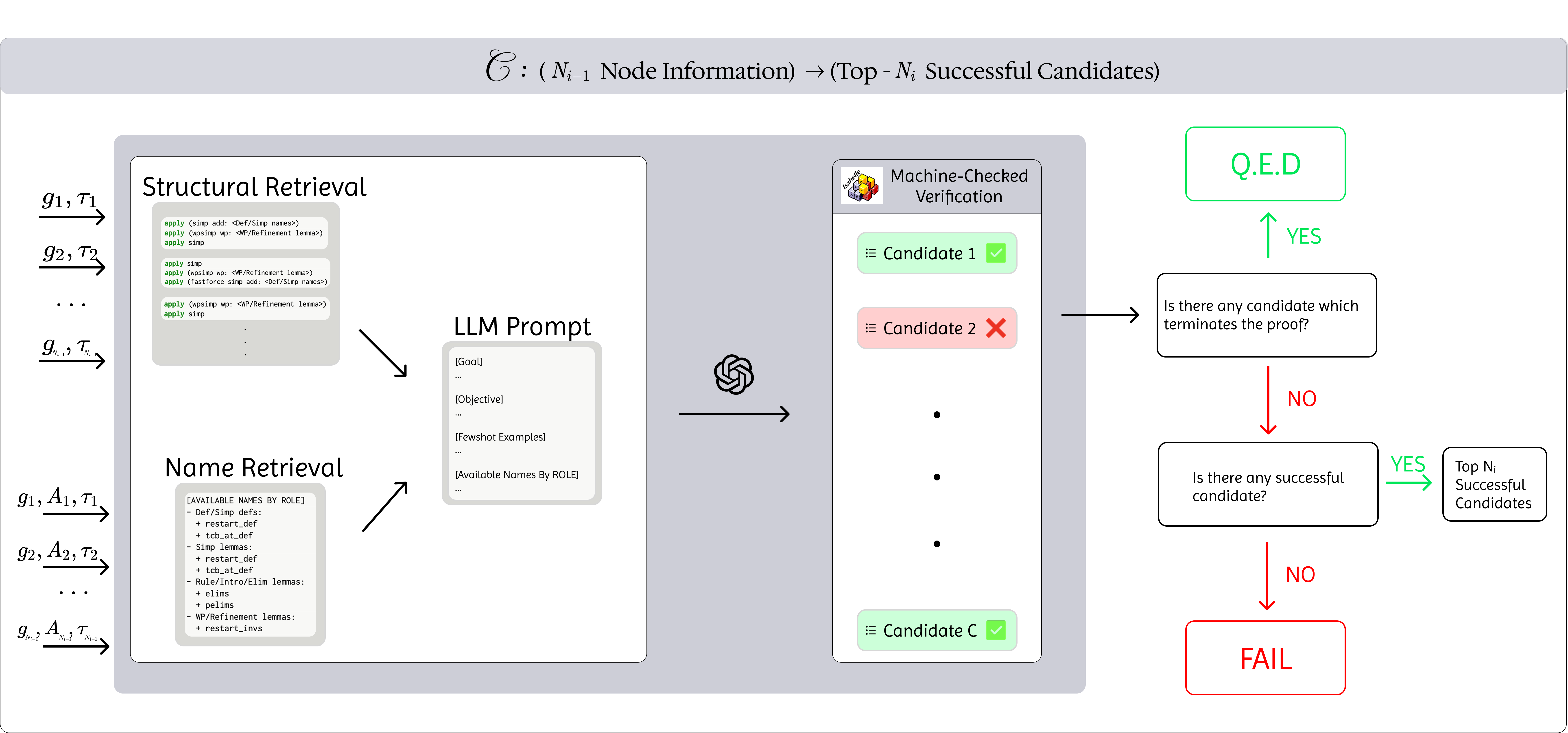}
    \caption{Command Generator  $\mathcal{C}$}
    \label{fig:module-A}
\end{figure}

Figure~\ref{fig:main-pipeline} illustrates the architecture of \sysname. Instead of generating a complete proof in a single pass, \sysname\ incrementally constructs proofs through short tactic-level continuations. PROMISE begins by obtaining the initial proof state of the target lemma, including the current goal and active assumptions, through the Scala–Isabelle bridge. This information defines the root proof node. \sysname\ then expands this node using the command generator $\mathcal{C}$ (Figure~\ref{fig:module-A}). Here, $\mathcal{C}$ denotes a subroutine that takes one or more proof nodes and returns up to $N_i$ proof commands for the next search depth-$i$. Applying the top-$B$ candidates returned by $\mathcal{C}$ to the root proof state yields $N_1$ depth-1 proof-state nodes. Here, $B$ denotes the beam width, and $N_1$ denotes the number of command candidates ($N_1 ≤ B$) that execute without error on the current proof state. At depth 1, \sysname\ passes all $N_1$ depth-1 proof-state nodes to $\mathcal{C}$  to obtain $N_2$ candidates. These candidates produce the depth-2 proof-state nodes. This process repeats until the search reaches the predefined depth bound $D$. If no candidate closes the proof by depth $D$, then \textsc{PROMISE} immediately terminates the whole proof generation pipeline and returns the message `[BEAM] failed to find proof within depth limit'. Otherwise, it attaches a proof terminating candidate to the proof prefix of the corresponding depth-$D$ proof state node to make a final proof script and terminates successfully. 
\begin{remark} A proof is accepted only when a candidate both closes the current proof state and passes a final whole-theory consistency check. \end{remark}

\subsection{Command Generator $\mathcal{C}$}
The command generator $\mathcal{C}$ takes one or more proof-state nodes from the current frontier and returns verified proof commands for expanding the next search depth. Algorithm~\ref{alg:promise-command-generator} illustrates what exactly happens inside the command generator $\mathcal{C}$ :

\begin{enumerate}[label={(\arabic*)}, leftmargin=2.5em, nosep]
    \item For each node in the current frontier, perform the following steps.
    \begin{enumerate}[label=\alph*., leftmargin=2em, nosep]
        \item Retrieve structurally related proof fragments and relevant theorem names,
        \item Construct a grounded few-shot prompt,
        \item Generate candidate proof steps using the language model,
        \item Normalize and statically filter the generated candidates,
        \item Verify candidates in Isabelle
    \end{enumerate}
    \item Collect all successful candidates produced from the current frontier.
    \item Select top-$B$ candidates among them.
    \begin{enumerate}[label=\alph*., leftmargin=2em, nosep]
        \item If there is no successful candidates, return \texttt{fail}.
        \item If any candidate closes the proof, return Q.E.D.
        \item If fewer than $B$ candidates succeed, return all successful candidates.
    \end{enumerate}
\end{enumerate}

\begin{remark}
If the regeneration limit is $R > 1$, then Steps~(1)--(3) can be repeated up to $R-1$ additional times.
\end{remark}

\begin{algorithm}[t]
\caption{Command Generator ($\mathcal{C}$)}
\label{alg:promise-command-generator}
\begin{algorithmic}[1]
\Require Current beam frontier $\mathcal{B}_i = \{n_1, \ldots, n_m\}$ with $1 \leq m \leq B$, beam width $B$, regeneration limit $R$, candidate budget $C$, target theorem $\tau$
\Ensure \textsc{Q.E.D.}, FAIL, or the next frontier of size at most $B$
\For{$r \gets 1$ to $R$}
    \State $\mathrm{Succ}_i \gets \emptyset$
    \ForAll{$n = (\pi, \sigma, A, g, k) \in \mathcal{B}_i$}
        \Comment{Context construction}
        \State $T \gets \textsc{StructuralRetrieval}(g,\tau)$
        \State $N \gets \textsc{NameRetrieval}(g,A,\tau)$

        \Comment{Prompting and candidate generation}
        \State $P \gets \textsc{BuildPrompt}(n,T,N)$
        \State $Cand \gets \textsc{GenerateCandidates}(P,C)$

        \Comment{Normalization and filtering}
        \State $Cand \gets \textsc{NormalizeAndFilter}(Cand,n,N,\tau)$
        
        \Comment{Machine-checked verification}
        \ForAll{$c$ in $Cand$}

            \State $res \gets \textsc{ExecuteInIsabelle}(n,c)$

            \If{\textsc{ProofClosed}$(res)$ \textbf{and} \textsc{FullTheoryCheck}$(\tau)$}
                \State \Return \textsc{Q.E.D.}
            \EndIf

            \If{\textsc{MakesProgress}$(res)$}
                \State $s \gets \textsc{BeamScore}(res)$
                \State $Succ_i \gets Succ_i \cup \{(\textsc{Node}(res), s)\}$
            \EndIf
        \EndFor
    \EndFor
    \If{$\mathrm{Succ}_i \neq \emptyset$}
        \State \Return top $N_i = \min(B, \lvert \mathrm{Succ}_i \rvert)$ candidates in $\mathrm{Succ}_i$
    \EndIf
\EndFor
\State \Return FAIL
\end{algorithmic}
\end{algorithm}

\subsection{Proof-State Search Model}

\sysname\ performs beam search over Isabelle proof states.
Each node in the beam represents a partially constructed proof together with its current proof context. Formally, a beam node is represented as
\[
n = (\pi, \sigma, A, g, k),
\]
where $\pi$ denotes the current proof prefix, $\sigma$ the Isabelle proof state, $A$ the active assumption set, $g$ the current goal text, and $k$ the number of remaining subgoals. The initial node is obtained by probing the target theorem in an isolated working theory using a Scala--Isabelle bridge. This probe produces the initial proof state together with its assumptions, goal text, and subgoal count. During search, candidate proof steps are appended to the proof prefix and executed in Isabelle to produce successor proof states.
Verified successors are ranked by a beam scoring function and retained to form the next beam. The proof search algorithm operating on this proof-state representation is described next.

\subsection{Core Proof Search Algorithm}

\begin{algorithm}[t]
\caption{\sysname: Retrieval-Grounded Beam Proof Search}
\label{alg:promise-core}
\small
\begin{algorithmic}[1]
\Require target theorem $\tau$, beam width $B$, candidate budget $C$, depth bound $D$
\Ensure valid proof of $\tau$ or FAIL

\State $(\sigma_0, A_0, g_0, k_0) \gets \textsc{ProbeInitialState}(\tau)$
\State $Beam \gets \{(\pi=\emptyset,\sigma_0,A_0,g_0,k_0)\}$

\For{$d = 1$ to $D$}

    \Comment{Call command generator at depth \texttt{d}}

    \If{$\mathcal{C}$ ($N_1, N_2, ..., N_{n_d}$) == FAIL}
        \State \Return{FAIL}
        
    \ElsIf {$\mathcal{C}$ ($N_1, N_2, ..., N_{n_d}$) == Q.E.D}
        \State \Return{Q.E.D}
        
    \Else
        \State \textbf{continue}

    \EndIf
    
\EndFor

\State \Return FAIL
\end{algorithmic}
\end{algorithm}

Algorithm~\ref{alg:promise-core} describes the core proof search procedure of \sysname. In fact, if you have a solid understanding of the Algorithm~\ref{alg:promise-command-generator}, this is actually a very simple algorithm. It iteratively invokes the command generator $\mathcal{C}$ at each search depth. During the loop, if $\mathcal{C}$ returns `FAIL', then it also returns `FAIL'. If $\mathcal{C}$ does not return `Q.E.D' for all depth, then \textsc{PROMISE} also returns `FAIL'. With the behavior of $\mathcal{C}$ defined, we can now describe the overall search procedure of \sysname. Algorithm~\ref{alg:promise-core} performs beam search over the proof states defined above. Starting from the initial proof state of the target theorem, \sysname\ iteratively expands the current beam. For each node, the system first constructs a retrieval-grounded context by identifying structurally related proofs and relevant theorem names. This context is used to build a grounded prompt for the language model, which then generates candidate proof continuations. The generated candidates are normalized and filtered to remove malformed syntax, invalid theorem references, and other implausible tactic expressions. The remaining candidates are executed in Isabelle through a Scala--Isabelle bridge to obtain successor proof states. Candidates that close the proof are accepted only if they also pass a final whole-theory consistency check. Otherwise, candidates that make measurable progress are inserted into a successor pool and ranked according to a beam scoring function. The highest-scoring successors form the beam for the next search iteration. The following subsections describe each stage of this pipeline in detail.

\subsection{Retrieval-Augmented Context Construction}

For candidate generation, \sysname\ augments each search node with retrieved contextual information. 
The framework performs two complementary retrieval operations: \emph{structural retrieval} and \emph{name retrieval}. 
The key idea of \sysname\ lies in structural retrieval, which identifies previously solved lemmas whose proof-state transition patterns resemble the current proof obligation. 
This enables the system to reuse recurring proof evolution patterns that arise across large verification developments. 
Name retrieval, in contrast, plays a supporting role by identifying concrete lemmas and definitions available in the current proof context. 
While structural retrieval captures how similar proofs progress, name retrieval determines which concrete facts can be safely used to construct the next proof transition.

The key idea of \sysname\ is \emph{goal-conditioned structural retrieval}.
Instead of retrieving artifacts based solely on the original theorem statement, 
the system retrieves previously solved lemmas whose proof-state transition patterns resemble the \emph{current} proof goal.
This allows the search procedure to reuse common proof evolution patterns that arise across large verification developments.

Structural retrieval consists of two phases: (1) Database (DB) construction, (2) few-shot example selection using the DB. In the first phase, we build a database containing structural representations of lemmas.
For each lemma included in the database, we execute its original proof commands line by line and record each intermediate goal together with the corresponding remaining proof-command suffix (proof commands which are not executed yet) into `\texttt{proof\_suffix.json}' during the process. We then extract the intermediate goal from each record, parse it with the official Isabelle parser, and embed the resulting representation into the database. At inference time, the resulting database is used to guide proof generation for the target lemma. The retrieval procedure is as follows :
\begin{enumerate}
    \item Parse the current goal with an official Isabelle parser,
    \item Retrieve top-30 scoring intermediate goal from DB,
    \item Rerank those with respect to semantic similarity with the target lemma,
    \item Choose top-8 reranked intermediate goal, 
    \item Get proof suffixes corresponding to those states from \texttt{proof\_suffix.json}, and
    \item Insert them as a fewshot examples in the prompt
\end{enumerate}
We next define the scoring metrics used for ranking and reranking.
Each retrieved lemma is first assigned a structural similarity score
\begin{equation}
s_{\mathrm{struct}}
=
(1-d)
+
\alpha \cdot \min(m, c_{\max})
+
\beta \cdot \ell,
\end{equation}
where $d$ denotes the embedding distance between the query and the candidate lemma, 
$m$ is the number of overlapping constants, 
$\ell$ measures the lexical overlap between the goals,
and $\alpha$ and $\beta$ are weighting constants.
Because structural similarity alone may retrieve proofs that are syntactically related but semantically irrelevant, 
the retrieved lemmas are further reranked with respect to the semantic similarity:
\begin{equation}
s_{\mathrm{rerank}}
=
0.55\,s_{\mathrm{struct}}
+
0.35\,s_{\mathrm{goal}}
+
0.10\,s_{\mathrm{const}},
\end{equation}
where $s_{\mathrm{goal}}$ measures similarity between the current goal and the retrieved lemma's initial goal, 
and $s_{\mathrm{const}}$ captures overlap in constants and salient symbols.
This second-stage ranking favors lemmas that are not only structurally similar but also semantically close to the target lemma. The following example illustrates how these two scores complement each other. During proof generation for the lemma \texttt{lookup\_cap\_and\_slot\_valid\_fault} using \textsc{PROMISE}, at the certain proof state, pure structural ranking would keep \texttt{not\_empty\_lc} in the top-8 retrieved lemmas: it took 3rd place by raw structural score with $s_{\mathrm{struct}} = 0.43.$
However, it was reranked with respect to the rerank score $s_{\mathrm{rerank}}$ by the expression (2).
Despite its moderate structural similarity $s_{\mathrm{struct}}$, \texttt{not\_empty\_lc} had relatively weak semantic relevance to the target lemma: its goal-similarity and constant-overlap terms were only
\[
s_{\mathrm{goal}} = 0.15, \qquad s_{\mathrm{const}} = 0.18,
\]
so its final score became
\[
0.55(0.43) + 0.35(0.15) + 0.10(0.18) = 0.31.
\]
As a result, \texttt{not\_empty\_lc} fell from third place in the pure structural ranking to 14 after semantic adjustment and was no longer retained in the top-8 few-shot pool.

By contrast, \texttt{get\_ipc\_buffer\_words} initially ranked much lower than \texttt{not\_empty\_lc} under pure structural retrieval: ranked 8th with $s_{\mathrm{struct}} = 0.34.$
But its semantic relevance to the target lemma was much higher than \texttt{not\_empty\_lc}, with
\[
s_{\mathrm{goal}} = 0.82, \qquad s_{\mathrm{const}} = 0.09,
\]
which gives
\[
s_{\mathrm{rerank}} = 0.55(0.34) + 0.35(0.82) + 0.10(0.09) = 0.48.
\]
This promoted \texttt{get\_ipc\_buffer\_words} from 8th under raw structural ranking to 3rd after semantic reranking. Similarly, \texttt{thread\_set\_tcb\_fault\_handler\_update\_invs} rose from rank 9 to rank 4. This example shows why reranking is necessary: a lemma that would have been selected by $s_{\mathrm{struct}}$ alone is displaced by lemmas with lower-structural similarity but stronger semantic alignment with the goal once $s_{\mathrm{goal}}$ and $s_{\mathrm{const}}$ are taken into account.

\sysname\ does not reuse the retained lemmas as full proofs.
Instead, \sysname\ extracts \emph{tactic templates} from them by preserving the overall proof shape while abstracting away the concrete theorem arguments.
These templates serve as few-shot exemplars that illustrate how similar proofs typically evolve, 
guiding the language model toward plausible proof-state transitions rather than encouraging direct proof copying.

\paragraph{\textbf{Name Retrieval}}

In addition to structural retrieval, \emph{name retrieval} determines which concrete facts are available for constructing the next proof step.
Its primary purpose is to ground candidate generation in the set of lemmas and definitions that are valid in the current proof context.

For each beam node, \sysname\ constructs a grounded vocabulary of theorem and definition names that are likely to be useful for the current proof state.
This vocabulary is assembled by aggregating candidates from four complementary sources:
\begin{enumerate}
    \item Implicit lemma names inferred from constants and identifiers appearing in the current goal,
    \item Character-$n$-gram retrieval over theorem names and lemma texts in the repository,
    \item Interactive theorem search within the live Isabelle context, and
    \item Verified definitional lemmas.
\end{enumerate}
Because lemmas ending in \texttt{\_def} are not explicitly defined in the codebase, \sysname\ cannot rely on a static whitelist for them. \sysname\ therefore proposes candidate \texttt{\_def} lemmas by appending \texttt{\_def} to constants or identifiers that appear in the current goal state. \textsc{PROMISE} then filters these candidates through Isabelle to remove unavailable lemmas. A second, more grounded source of \texttt{\_def} names is Proof-IDE (PIDE), which provides \texttt{\_def} lemmas available in the current proof context. Combining these two sources yields a set of plausible \texttt{\_def} lemma candidates.

The role of name retrieval differs from that of structural retrieval.
While structural retrieval captures how similar proofs typically evolve by identifying reusable proof-state transition patterns,
name retrieval identifies which concrete facts can be safely used in generated commands.
The retrieved names are organized into semantic role buckets, including definitions, simplification lemmas, rule-style lemmas, and WP/refinement lemmas. These categories are used to ground prompt construction and constrain candidate validity.

\paragraph{\textbf{Verification of Definitional Lemmas}}

To ensure that unfolding hints used during generation are valid in the current proof context, \sysname\ explicitly verifies candidate definitional lemmas with Isabelle. 
Definitional lemmas, particularly those ending in \texttt{\_def}, cannot always be reliably recovered from static repository text alone. 
Therefore, \sysname\ queries Isabelle directly to check whether each candidate definitional lemma is available in the active proof environment. 
Only lemmas confirmed by Isabelle are retained, ensuring that all unfolding hints supplied to the language model are context-valid rather than merely textually plausible.

\paragraph{\textbf{Leakage Prevention}}

To prevent proof leakage, \sysname\ explicitly excludes the target theorem from all retrieval stages. 
Structural retrieval omits the target theorem by name, and name-based retrieval results are filtered to remove any references to it. 
A final safety pass further eliminates any remaining self-references. 
As a result, the original proof of the target theorem is never used as a retrieval exemplar or a hint source during search.

\subsection{Grounded Few-Shot Prompting}

From the current beam node and the retrieved context, \sysname\ constructs a grounded few-shot prompt for next-step generation. The prompt includes the current goal, active assumptions, the proof prefix accumulated so far, compact feedback summarizing recent failed attempts, structurally retrieved tactic templates, and the retrieved theorem names organized by role.

A key component of the prompt is method diversification. \sysname\ estimates which tactic families are promising for the current state by combining three signals: the current goal family, the methods observed in retrieved tactic templates, and persistent coverage statistics accumulated from previously successful state transitions. 
Methods are ranked lexicographically according to goal-family compatibility, local evidence from retrieved templates, global rarity, and repository-level prevalence. From this ranking, the system derives (i) \emph{method targets}, which specify desired coverage across tactic families, and (ii) \emph{candidate method slots}, which assign preferred initial methods to specific candidate positions. This discourages collapse into a single overused method family and encourages a controlled diversity of candidate tactics.

The model is prompted to generate a fixed number of short candidate next steps, typically one or two Isabelle commands per candidate, subject to the grounded facts and method preferences encoded in the prompt.

\subsection{Candidate Normalization and Static Filtering}

To guarantee syntactic validity and contextual grounding before machine-checked verification,
\sysname\ applies a deterministic normalization and filtering stage prior to verification. 
The raw language-model outputs are first transformed into a canonical candidate list through a repair process that standardizes syntax, resolves aliases, removes malformed fragments, and replaces weakly grounded theorem references with retrieved context-valid names when possible.

Nevertheless, the generated outputs may occasionally be weak, for example, when they are highly repetitive, malformed, or insufficiently diverse. 
In such cases, the system augments the candidate pool with grounded fallback candidates constructed directly from retrieved theorem names and tactic schemas. 
These fallback candidates are assembled from validated facts and predefined tactic patterns, such as simplification, rule application, or WP-style reasoning. 
This mechanism ensures that the candidate pool remains grounded even when the generative output is unreliable.

After normalization, each candidate is subjected to deterministic static checks.
Candidates are rejected if they are empty, duplicated, syntactically implausible, dependent on unknown methods, excessively timeout-prone, reliant on unresolved tactic variables, or reference facts outside the grounded allowed set. 
Candidates that reference the target theorem itself are also discarded. 
These checks eliminate obvious failure modes before Isabelle execution and significantly reduce wasted verification effort.

\subsection{Machine-Checked Verification}

To guarantee the semantic correctness of generated proof steps, \sysname\ verifies all surviving candidates through machine-checked execution in Isabelle. 
Each candidate is appended to the parent proof prefix and executed via the Scala--Isabelle bridge to probe the resulting proof state. 
For each execution, the verifier obtains the successor proof state, successor assumptions, successor goal, and the number of remaining subgoals. 
Verification results are memoized using a normalized task signature so that duplicate state transitions are not re-executed.

A candidate execution may yield one of three outcomes. 
First, the candidate may fail, in which case it is discarded and its failure signal is summarized into compact error feedback for future regeneration. 
Second, the candidate may partially succeed by transforming the proof state without completing the proof; in this case, it becomes a scored successor node for further search. 
Third, the candidate may appear to close the proof locally. 
Even in this case, \sysname\ does not accept the proof immediately. 
Instead, the system reconstructs the full proof in the isolated theory and requires the entire Isabelle build to succeed. 
This final check ensures that locally closed proofs are also globally valid within the full theory environment.

\subsection{Beam Scoring and Adaptive Update}

To prioritize promising proof states during search, \sysname\ assigns a beam score to each verified successor based on proof progress, proof compactness, and tactic diversification. 
Let $k_{\mathrm{parent}}$ and $k_{\mathrm{child}}$ denote the parent and successor subgoal counts, respectively. 
The immediate progress gain is defined as
\begin{equation}
\Delta = \max(0, k_{\mathrm{parent}} - k_{\mathrm{child}}).
\end{equation}

The beam score of a successor node is computed as
\begin{equation}
s_{\mathrm{beam}}
=
-k_{\mathrm{child}}
-0.01L
+0.25\Delta
+b(m),
\end{equation}
where $L$ is the resulting proof-prefix length and $b(m)$ is an optional diversification reward associated with the first method $m$ used by the candidate. 
This reward is defined as
\begin{equation}
b(m)
=
\min\!\left(
\gamma_{\mathrm{cap}},
\frac{\gamma_w}{1+\sqrt{u(m)}}
\right),
\end{equation}
where $u(m)$ is the accumulated usage count of method $m$, $\gamma_w$ is a reward weight, and $\gamma_{\mathrm{cap}}$ is a reward cap. 
This scoring formulation favors candidates that reduce subgoals, maintain concise proofs, and explore less-saturated tactic families.

After scoring, all verified successor nodes from the current frontier are pooled and ranked. 
The next beam is formed by selecting the highest-scoring successors up to the active beam width. 
The beam width and candidate budget may be adjusted dynamically in response to timeout pressure and observed progress. 
When the nominal depth bound is reached but promising frontier states remain, the search depth may be extended up to a predefined hard cap.

\subsection{Termination}

\sysname\ terminates successfully only when a candidate sequence both closes the local proof state and passes the final whole-theory Isabelle check. Otherwise, the search terminates unsuccessfully after exhausting the beam horizon, regeneration budget, or adaptive search limits. Overall, \sysname\ is therefore a retrieval-grounded, machine-checked, beam-search framework for incremental theorem proving, rather than a single-shot language-model decoding procedure.

\section{Evaluation Setup}\label{sec:evaluation-setup}

\paragraph{\textbf{Benchmark and Isabelle/HOL setup.}}
We evaluate \sysname\ on a benchmark of 223 Isabelle/HOL theorems achieved from the seL4 \texttt{l4v} development. The whole benchmark is defined in \texttt{benchmark.json} and is evenly split into 100 \texttt{P1} tasks and 100 \texttt{P2} tasks, excluding 23 \texttt{P3} lemmas. The theorems come from both library and proof-oriented developments, including \texttt{lib/Monads}, \texttt{proof/invariant-abstract}, \texttt{proof/infoflow}, \texttt{proof/sep-capDL}, and \texttt{proof/access-control}. This choice of source theories is deliberate: it is intended to cover a broad but coherent range of reasoning patterns within the Isabelle/HOL verification ecosystem. \texttt{lib/Monads} contributes generic monadic definitions, weakest-precondition rules, and reusable proof infrastructure that appears throughout the seL4 proofs. \texttt{proof/invariant-abstract} captures the core invariant-preservation arguments over the seL4 abstract specification, while \texttt{proof/access-control} and \texttt{proof/infoflow} add relational and security-oriented reasoning about integrity, authority confinement, confidentiality, and noninterference. \texttt{proof/sep-capDL} contributes a different proof style based on separation logic over the capDL model, including frame-style and heap-decomposition arguments. As a result, the benchmark is neither restricted to generic library lemmas nor confined to one narrow verification layer: it evaluates whether a prover can handle reusable algebraic reasoning, monadic Hoare-style proofs, system-specific invariant arguments, security proofs, and separation-logic reasoning within the same maintained corpus.
For each benchmark instance, we isolate the target theorem into a temporary working theory, construct fresh dependency and target sessions, and interact with Isabelle through a Scala–Isabelle bridge. A theorem is counted as solved only if the synthesized proof both closes the local proof state and passes a final whole-theory Isabelle build. This evaluation protocol avoids false positives and ensures that all reported successes correspond to replayable Isabelle/HOL proofs.

\paragraph{\textbf{Why seL4?}}
\label{sec:benchmark-sel4}
Our goal is to automate formal verification for large-scale systems, and recent studies have made such attempts on a few large-scale systems. Among them, We decide to use seL4/l4v as a benchmark because it includes interactive proof datasets that are both large in scale and technically very challenging, yet hold significant practical importance in real-world systems. The seL4 project established the first machine-checked functional-correctness proof for a complete general-purpose operating-system kernel, connecting an abstract specification to the kernel's C implementation \cite{Klein2009seL4}. Subsequent work extended this result into a broader proof stack that includes binary correctness, access-control enforcement, information-flow security, and other end-to-end assurance arguments \cite{Klein2014Comprehensive,Murray2013InfoFlow,Klein2017Provably}. In sum, l4v is not a collection of isolated textbook lemmas: it is a decent verification system spanning abstract specifications, executable models, refinement proofs, C-level reasoning, and security proofs.
This makes l4v particularly suitable for evaluating proof generation performance of large language models. Good performance on this corpus requires substantially more than local tactic imitation. A prover must handle long-range dependencies, premise selection in a large theory graph, proof-state-sensitive tactic generation, and reasoning across multiple abstraction levels. These demands align closely with the intended capabilities of retrieval-augmented and state-aware LLM proving systems. The benchmark is also structurally diverse: the repository exposes specifications, refinement layers, access-control and information-flow proofs, binary-verification components, and proof-engineering infrastructure within a single Isabelle/HOL development \cite{l4vRepo}. In the snapshot used in this work, a simple scan of \texttt{verification/l4v} found 2{,}016 Isabelle theory files, approximately 66{,}913 theorem-style declarations, and 14{,}776 definition-style declarations. For our intermediate-goal retrieval experiments, replay over this snapshot yielded 72{,}749 indexed proof states.
A further advantage is that seL4/l4v is an actively maintained, open-source project rather than a static benchmark assembled solely for theorem proving. The project positions seL4 as a high-assurance foundation for safety- and security-critical systems, and official materials emphasize the continuous maintenance of its proofs as the implementation evolves \cite{sel4WhitePaper2026,sel4Verification2026,sel4FactSheet2026}. This plays an important role in selecting benchmark for LLM-based proof generation: a useful benchmark should reflect the realities of real-world proof engineering, where generated proofs must remain valid within a large, continually updating codebase. For these reasons, seL4/l4v provides a demanding, realistic, and reliable benchmark for research on automated proof generation with language models.

\paragraph{\textbf{Models and hardware.}}
All experiments are run on Ubuntu 24.04.2 LTS on a server with 96 logical cores, 503\,GiB of RAM, and four NVIDIA RTX 6000 Ada GPUs (49\,GB each). We evaluate \textsc{PROMISE} with multiple language-model backends: \texttt{Qwen2.5-Coder-7B-Instruct}, \texttt{GPT-3.5-turbo}, and \texttt{GPT-4.1}. We used \texttt{GPT-3.5-turbo} and \texttt{GPT-4.1} to compare \sysname\ performance with Selene. Additionally, in order to establish the scale-invariance of our framework, we incorporated \texttt{Qwen2.5-Coder-7B-Instruct} as a leaner alternative to the larger ones we used (\texttt{GPT-3.5-turbo}, \texttt{GPT-4.1}). Unless otherwise stated, all systems in a given comparison use the same underlying model backend.

\paragraph{\textbf{Compared systems and re-implementations.}}
We compare \textsc{PROMISE} against two representative retrieval-augmented theorem-proving baselines, Selene and Rango. Since the original Selene artifact is not publicly available, we implemented our own version following the algorithm and evaluation protocol described in the paper. In our experiments, the behavior of this re-implementation with \texttt{GPT-4.1} closely matches the trends reported in the original Selene paper, which provides evidence that the implementation is faithful enough for comparison. In contrast, the original Rango artifact is publicly available, but it targets Rocq/Coq rather than Isabelle/HOL. We therefore re-implemented an Isabelle/HOL version of Rango by following its published retrieval, prompting, and search procedure as closely as possible within the Isabelle setting. The re-implemented versions may not fully match the original ones. However, for convenience, we will refer to these re-implemented versions as the original model name in this paper. 

\begin{table}[th]
\centering
\begin{tabular}{lcccc}
\hline
\multicolumn{5}{c}{\textbf{Qwen2.5-Coder-7B-Instruct}} \\
\hline
Method & Selene ACC5 & Selene ACC30 & Increase & PROMISE\\
\hline
P1 & 30.0\% & 40.0\% & +10.0 pp & 77\%\\
P2 & 2.0\%  & 5.0\%  & +3.0 pp & 36\%\\
P3 & 8.7\%  & 13.0\%  & +4.3 pp & 30.4\%\\
\hline
\end{tabular}
\caption{Accuracy comparison between Selene ACC5, Selene ACC30, and PROMISE on Qwen2.5-Coder-7B-Instruct.}
\label{tab:selene-30}
\end{table}

\paragraph{\textbf{Baseline configurations.}}
For Selene, we use the standard \textsc{ACC\#1} and \textsc{ACC\#5} protocols: one attempt with temperature 0.0 and five attempts with temperature 0.5, respectively. Temperature, which ranges from 0 to 1, controls the randomness of LLM responses: values close to 0 lead to more conservative and deterministic responses, while values closer to 1 produce more diverse and less predictable outputs. Hence such adjustment enables us to evaluate LLM's proof generation performance across different levels of creativity.
In both cases, we use five demonstrations, top-$p=0.95$, a maximum output length of 2,048 tokens, and a verification timeout of 600 seconds. We use a verification timeout of 600 seconds to balance fairness and practicality. This budget is long enough to accommodate nontrivial Isabelle replay and automation costs on the harder \texttt{l4v} theorems, but also short enough to keep evaluation bounded and to avoid letting a few pathological instances dominate runtime. For Rango, we use rollout search with 20 rollouts and a maximum of 24 steps per rollout, BM25 proof retrieval with top-$k=8$, TF--IDF lemma retrieval with top-$k=12$, and temperature 1.0. The original Rango evaluation is primarily controlled by a per-lemma timeout. In our Isabelle/HOL setting, however, repeated interaction with Isabelle through the Scala--Isabelle bridge incurs substantial overhead for retrieving proof states and probing candidate steps. Under a pure wall-clock budget, some lemmas terminate before enough rollout steps are explored, even though successful proof trajectories may still exist. We therefore use a fixed rollout-step budget, capping each rollout at 24 steps, to reduce the distortion introduced by bridge overhead and to obtain a more stable comparison across lemmas. This fixed budget is conservative relative to \sysname, which issues about 30 LLM queries per lemma on average on the same benchmark. For Selene, Table~\ref{tab:selene-30} demonstrates that increasing retrial budget does not affect success rate dramatically as the prompt remains the same regardless of the number of queries. In addition, the Table~\ref{tab:selene-30} also shows that Selene's improved accuracy is much lower than the success rate of \sysname. Therefore, we just used the Selene's default retrial limit (ACC1, ACC5) in our evaluation. 

\paragraph{\textbf{\textsc{PROMISE} configuration.}}
Unless otherwise stated, \sysname\ uses the default retrieval-grounded beam-search configuration described in Section~\ref{sec:methodology}. The search starts with beam width 6 and regeneration limit 2. The initial depth bound is 10, and the search may extend this bound by one step at a time up to a hard cap of 12 when promising successor states remain. Candidate generation uses temperature 0.9. Moreover, \sysname\ starts with 12 candidates per state. Candidate verification uses a 120-second timeout per machine-checked probe. During search, \sysname\ adaptively adjusts both beam width and candidate budget according to timeout pressure and observed proof progress.

\paragraph{\textbf{Prompt Design.}}
Figure~\ref{fig:promise-prompt} is a simplified \sysname\ prompt example. \sysname\ uses a single, high-density prompt that provides the language model with a complete snapshot of the current proof state, including current goal state, active assumptions, the partial proof prefix, and verifier feedback for the previously probed proof commands. In order to facilitate structural information, the prompt contains up to five proof templates retrieved via structural retrieval. We deliberately hide lemma names used in those templates in order to clearly separate structural retrieval and name retrieval pipelines. Moreover, to make the generative process reliable as much as possible, the prompt provides a role-partitioned theorem inventory. This inventory categorizes accessible premises into functional groups-definitions, simplification facts, introduction/elimination rules, and WP/refinement lemmas. Finally, the model is asked to synthesize diverse candidates of tactic sequences by following the human-like proof reasoning principle we gave to it. 
\begin{remark} As finishing the proof in a single line is not feasible for challenging lemmas in most cases and is not consistent with our iterative proof mining approach, we forced the language model to use 'apply' command instead of using 'by' command, which is used for single line proofs. \end{remark}

\begin{figure}[h]
\centering
\setlength{\fboxsep}{5pt}
\fbox{%
\begin{minipage}{0.94\columnwidth}
\footnotesize
\setlength{\tabcolsep}{3pt}
\renewcommand{\arraystretch}{1.05}
\raggedright
\begin{tabular}{@{}p{0.22\linewidth}p{0.74\linewidth}@{}}
\textbf{System role.} & Isabelle/HOL proof engineer for seL4/l4v.\\[-0.25ex]
\end{tabular}
\begin{tabular}{@{}p{0.22\linewidth}p{0.74\linewidth}@{}}
\textbf{Task.} & Generate exactly 8 diverse next-step candidates (1--2 commands each) for the current checked goal. Prefer retrieval-grounded tactics and use \texttt{apply}-style commands only.
\\[0.5ex]
\end{tabular}
\hrule
\vspace{0.4ex}
\textbf{Current state.}\\[-0.6ex] 
\vspace{0.8ex}
\begin{tabular}{@{}p{0.22\linewidth}p{0.74\linewidth}@{}}
- Goal & \parbox[t]{\linewidth}{%
\ttfamily
\char123 invs and (\(\lambda\)s. e \(\neq\) Interrupt \(\longrightarrow\) ct\_running s)\char125\newline
call\_kernel e\newline
\char123 \(\lambda\)rv. invs and (\(\lambda\)s. ct\_running s \(\vee\) ct\_idle s)\char125%
} \\

- Assumptions & none \\
- Proof prefix & none \\
- Feedback & avoid context-local pseudo-facts such as \texttt{assms}, \texttt{this}, \texttt{that}, and \texttt{thesis} \\
\end{tabular}
\vspace{0.4ex}
\hrule
\vspace{0.4ex}
\textbf{Structural retrieval templates.}\\[-0.6ex] 
\vspace{0.8ex}
\begin{tabular}{@{}p{0.22\linewidth}p{0.74\linewidth}@{}}
- T1 & \texttt{apply simp} $\rightarrow$ \texttt{apply (wpsimp wp: <WP/refinement lemma>)} \\
- T2 & \texttt{apply (cases x)} $\rightarrow$ \texttt{apply (simp add: <Def/Simp names>)} \\
- T3 & \texttt{apply (rule <Rule lemma>)} $\rightarrow$ \texttt{apply simp} \\
\end{tabular}
\vspace{0.4ex}
\hrule
\vspace{0.4ex}
\textbf{Role-partitioned theorem inventory.}\\[-0.6ex] 
\vspace{0.8ex}
\begin{tabular}{@{}p{0.22\linewidth}p{0.74\linewidth}@{}}
- Defs/Simp & \texttt{invs\_def}, \texttt{call\_kernel\_def}, \texttt{and\_def}, \dots \\
- Rules & \texttt{intro}, \texttt{spec}, \texttt{allE}, \texttt{conjE}, \dots \\
- WP/Ref. & \texttt{activate\_invs}, \texttt{do\_user\_op\_invs}, \texttt{kernel\_entry\_invs}, \texttt{timer\_tick\_invs}, \dots \\
\end{tabular}
\vspace{0.4ex}
\hrule
\vspace{0.4ex}
\textbf{Generation contract.}\\[-0.6ex] 
\vspace{0.8ex}
\begin{tabular}{@{}p{0.22\linewidth}p{0.74\linewidth}@{}}
- Grounding & use only names from the theorem inventory; never invent lemmas or copy hidden template names \\
- Reasoning & follow structure first, then instantiate grounded names, and favor progress-- making next steps over one-shot closure \\
- Diversity & cover multiple tactic families (\texttt{simp}, \texttt{rule}, \texttt{wp}, \texttt{struct}) and avoid near-duplicates \\
- Output & return exactly 8 short \texttt{apply}-style candidates, each a plausible next proof step \\
\end{tabular}
\end{minipage}%
}
\caption{Simplified \sysname\ prompt example.}
\label{fig:promise-prompt}
\end{figure}

\section{Evaluation}\label{sec:evaluation}

\begin{table}[t]
\centering
\setlength{\tabcolsep}{7pt}
\begin{tabularx}{\textwidth}{l *{7}{Y}}
\hline
& \multicolumn{3}{c}{\textbf{Qwen2.5-Coder-7B-Instruct}} & \multicolumn{2}{c}{\textbf{GPT-3.5-turbo}} & \multicolumn{2}{c}{\textbf{GPT-4.1}}\\
\cline{2-8}
\textbf{Method} & \textbf{P1} & \textbf{P2} & \textbf{P3} & \textbf{P1} & \textbf{P2} & \textbf{P1} & \textbf{P2}\\
\hline
\textbf{Selene ACC1} & 22 & 2 & 2 & 34 & 9 & 46 & 15\\
\textbf{Selene ACC5} & 30 & 2 & 2 & 42 & 14 & 54 & 23\\
\textbf{Rango} & 57 & 21 & 3 & 35 & 14 & 47 & \textbf{35}\\
\textbf{PROMISE} & \textbf{77} & \textbf{36} & \textbf{7} & \textbf{85} & \textbf{40} & \textbf{69} & 33\\
\hline
\end{tabularx}
\caption{Evaluation results on the full benchmark, grouped by proof level.}
\label{tab:eval-results}
\end{table}

Table~\ref{tab:eval-results} shows our benchmark results. As each proof level contains 100 problems except for \texttt{P3}, the raw counts in the table directly deduce success rates. In order to ensure practical scalability and localizability for complex proof engineering, we evaluate \texttt{P3} lemmas exclusively using Qwen2.5-Coder-7B-Instruct, rather than larger proprietary models like GPT-3.5-turbo or GPT4.1. Notably, \sysname\ demonstrates superior accuracy on \texttt{P3}, correctly proving 7 lemmas-more than double the performance of the best baseline, Rango (3). For \texttt{P1}/\texttt{P2}, \sysname\ achieves 77/36 on Qwen2.5-Coder-7B-Instruct, 85/40 on GPT-3.5-turbo, and 69/33 on GPT-4.1, corresponding to overall success rates of 56.5\%, 62.5\%, and 51.0\%, respectively. Averaged across all seven model-level settings, \sysname\ reaches 55.7\%, substantially above Rango (34.0\%), Selene ACC5 (26.8\%), and Selene ACC1 (20.9\%). Importantly, PROMISE obtains the best result in six of the seven evaluated settings, with the only exception being GPT-4.1 on P2, where Rango remains slightly higher (35 vs.~33).

\paragraph{\textbf{Why \sysname\ Does Not Surpass Rango on GPT-4.1/P2.}}
The only case in which \sysname\ does not outperform Rango is GPT-4.1 on P2. The difference is 2\%. We conjecture that this exception arises from the interaction between model capability and prompt strictness. GPT-4.1 is substantially stronger than earlier models at instruction following, long-context comprehension, distractor filtering, and multi-hop reasoning \cite{openai-2025-gpt41}. Recent evidence further suggests that, for strong models, few-shot chain-of-thought prompting does not necessarily improve reasoning: zero-shot CoT can match or exceed few-shot CoT, and the model often focuses primarily on the instruction rather than learning additional reasoning skill from the exemplars \cite{cheng-etal-2025-revisiting-chain}. More generally, analyses of CoT prompting show that intermediate reasoning steps often function mainly to communicate task format or task understanding, rather than to teach a genuinely better solution procedure \cite{madaan-etal-2023-makes}; other studies likewise find that CoT prompting can be beneficial in some sequential reasoning settings but detrimental in others, and that adding more in-context examples does not consistently help \cite{bellos-etal-2024-large}. Related prompt studies also report that prompt format can matter more than detailed descriptive instructions \cite{tang-etal-2025-large-language}. Taken together, these findings suggest that PROMISE's beam-fewshot prompt may have overconstrained GPT-4.1 on hard P2 goals by prescribing a narrower reasoning protocol than the model would otherwise choose. Because GPT-4.1 is especially strong at following nuanced instructions, it may have adhered more faithfully to this less efficient scaffold instead of exploiting its own stronger latent reasoning strategy. Rango's prompt, being less prescriptive while still providing iterative proof-state guidance, likely leaves more room for GPT-4.1 to choose an effective search trajectory, which plausibly explains why Rango retains a slight advantage only in this setting.

\paragraph{\textbf{Comparison with Selene.}}
\sysname\ consistently outperforms Selene, including Selene's stronger ACC5 configuration, across all models and both proof levels. The absolute gains over Selene ACC5 are +47 and +34 points on Qwen2.5-Coder-7B-Instruct, +43 and +26 on GPT-3.5-turbo, and +15 and +10 on GPT-4.1 for P1 and P2, respectively. This gap is notable because Selene ACC5 already allows multiple attempts, whereas Selene ACC1 corresponds to a single attempt. The relatively modest improvement from ACC1 to ACC5 suggests that repeated single-shot generation alone is insufficient; the larger gains of PROMISE instead come from retrieval-guided proof search.

\paragraph{\textbf{Stability across LLMs.}}
\sysname\ also exhibits stable behavior across model families. On P1, its success rate remains between 69\% and 85\%, and on P2 between 33\% and 40\%. The overall spread across Qwen2.5-Coder-7B-Instruct, GPT-3.5-turbo, and GPT-4.1 is 11.5 points, which is smaller than the variation observed for Rango and Selene. This result suggests that \sysname\ does not depend on a particular LLM family to be effective; rather, the retrieval-and-mining strategy transfers well across both open-weight and proprietary models.

\paragraph{\textbf{Keyword retrieval vs.~structural retrieval, and single-shot vs.~iterative mining.}}
The comparison with Selene and Rango helps isolate the source of the improvement. Selene represents a single-shot setting with repeated attempts (ACC1/ACC5), while Rango performs iterative mining with keyword-based retrieval over proof states and lemmas. PROMISE further strengthens iterative mining with two complementary retrieval channels: structural retrieval, which surfaces proofs with analogous shapes and reusable tactic patterns, and name retrieval, which provides concrete lemma and definition names needed to instantiate those patterns. This combination yields clear gains over keyword-based iterative mining in five of the six settings: compared with Rango, PROMISE improves by +20/+15 on Qwen2.5-Coder-7B-Instruct, +50/+26 on GPT-3.5-turbo, and +22 on GPT-4.1/P1, while trailing by only 2 points on GPT-4.1/P2. Overall, these results indicate that structural retrieval and name retrieval are both effective, and that their benefit is amplified when used inside an iterative mining loop rather than in a purely single-shot pipeline.

\begin{figure}
\centering
\includegraphics[width=\textwidth]{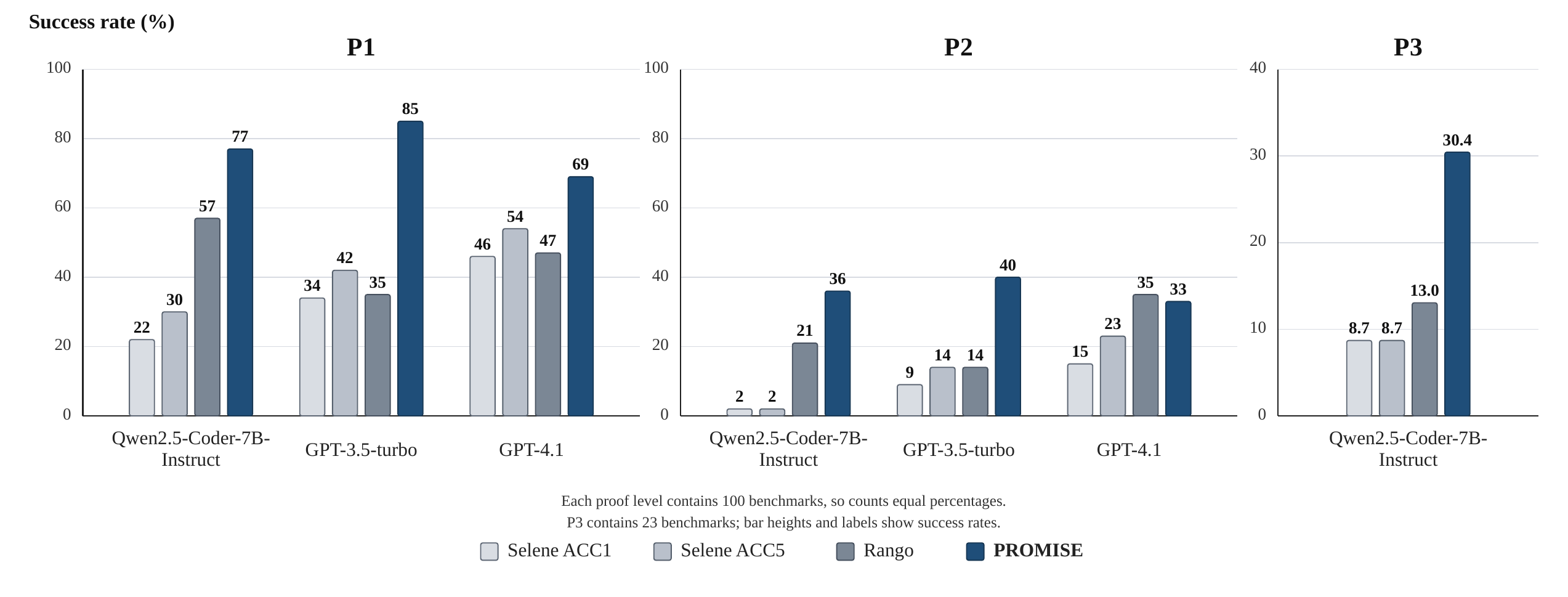}
\caption{Evaluation Result}
\label{fig:result}
\end{figure}

\section{Related Works}\label{sec:related-works}

Recent work on LLM-assisted theorem proving and software verification has rapidly expanded across multiple proof assistants, including Rocq/Coq, Lean, and Isabelle. Existing efforts can be broadly grouped into several directions. Those directions include large-scale benchmarks and environments, logic- and search-based automation, retrieval-augmented proving, tactic \& proof abstraction, and decomposition- or agent-based reasoning. Our work is most closely related to retrieval-augmented proving, but differs in a crucial way. Prior approaches typically retrieve artifacts such as premises, lemmas, whole proofs, or nearby textual context, whereas we focus on retrieving structural reasoning units that better reflect how proof knowledge is reused across large proof developments.

\paragraph{\textbf{Large-scale benchmarks}.}

A first line of work has developed datasets and environments for evaluating LLMs on formal reasoning. LeanDojo~\cite{NEURIPS2023_44414694} provides a retrieval-augmented environment for Lean and established the premise-selection-plus-tactic-generation paradigm for theorem proving. Selene~\cite{zhang-etal-2024-selene} introduced a project-level benchmark for industrial software verification based on seL4~\cite{Klein_EHACDEEKNSTW_09}, emphasizing complex inter-file dependencies, isolated verification environments, and whole-proof generation in Isabelle. Rango contributed CoqStoq~\cite{10.1109/ICSE55347.2025.00161}, a large Rocq/Coq corpus containing 2,226 repositories and 196,929 theorems, together with a benchmark targeting well-maintained real-world projects such as CompCert~\cite{Leroy-backend}. More recent work has extended evaluation beyond mathematical reasoning to system-software verification, for example using FSCQ~\cite{chen2015fscq} as a benchmark for proof search over a realistic verified file system. Collectively, these efforts demonstrate that large-scale proof developments differ substantially from small mathematical benchmarks: they exhibit deep dependency graphs, strong context sensitivity, and extensive intra-project reuse.

\paragraph{\textbf{Logic- and search-based proof automation}.}

Before the emergence of LLM-assisted proving, substantial progress had been made in automating proofs through logic-based and search-based techniques. In Isabelle, Sledgehammer~\cite{DBLP:conf/cade/BohmeN10} integrates automated theorem provers and SMT solvers with the interactive proof environment, automatically selecting premises and reconstructing proofs inside Isabelle. CoqHammer~\cite{DBLP:journals/jar/CzajkaK18} brings a similar architecture to Rocq/Coq by combining premise selection, automated theorem prover based proof search, and proof reconstruction. These systems have proven highly effective at discharging many proof obligations automatically.
More broadly, SMT-based verification frameworks such as Verus~\cite{10.1145/3694715.3695952} and Dafny~\cite{10.5555/1939141.1939161} leverage automated solvers to reason about program properties encoded as verification conditions. Once a program specification is translated into a suitable logical fragment, these systems can automatically discharge large classes of verification obligations.
However, these approaches primarily rely on logical encodings and solver-driven proof search rather than reuse of knowledge from existing proof corpora. Their automation strength depends mainly on the expressiveness of the underlying logic and the capabilities of external solvers. In contrast, LLM-assisted approaches try to leverage reusable proof knowledge embedded in large proof developments themselves.

\paragraph{\textbf{Retrieval-augmented proving}.}

A large body of work augments proof generation with retrieved context. LeanDojo~\cite{NEURIPS2023_44414694} retrieves relevant premises for Lean proofs. Rango~\cite{10.1109/ICSE55347.2025.00161} extends this idea in Rocq/Coq by retrieving both relevant lemmas and similar prior proofs from the current project and updating retrieval dynamically as the proof state evolves. Empirically, Rango shows that retrieving similar proofs—not just lemmas—substantially improves synthesis performance; its proof retriever alone contributes a large fraction of the gain over non-retrieval variants. In FSCQ-based proof search, additional project-level context likewise improves proof coverage.
These systems clearly demonstrate that retrieval is valuable for large proof developments. However, the retrieved objects remain primarily artifact-level units such as premises, proof scripts, theorem statements, or nearby textual context. As a result, they often provide context that is semantically relevant but structurally too coarse to capture the precise reasoning step required at the current proof state.

\paragraph{\textbf{Whole-proof generation, repair, and agentic verification}.}

Others treat proof automation more holistically. Baldur~\cite{Baldur} studies whole-proof generation/repair with LLMs. Selene similarly evaluates end-to-end proof generation and iterative fixing using verifier feedback in the seL4 setting. Recent agentic systems push this further. An experience report on CertiCoq~\cite{10.1145/3473591} shows that an LLM-assisted workflow produced an approximately 7,800-line Rocq proof for  administrative normal form correctness by adapting an existing continuation-passing style proof template, illustrating the growing feasibility of large verified developments under human guidance.  These systems are compelling, but their reuse mechanism is still usually framed at the level of full proof templates, surrounding context, or task-specific scaffolds, rather than reusable structural proof patterns.

\paragraph{\textbf{Tactic abstraction and proof compression}.}

A closely related direction seeks reusable abstractions below the whole-proof level. 
TacMiner~\cite{10.5281/zenodo.15761151} discovers tactic libraries from existing Rocq proofs using tactic dependence graphs, learning more tactics than a prior baseline, reducing proof corpus size by 26\%, and improving Copra~\cite{10.1145/2851581.2892333}’s proof automation success rate from 22\% to 60\%. 
This work highlights that reusable proof knowledge often lies in recurring tactic patterns rather than in complete proof scripts. 
However, tactic discovery primarily focuses on constructing re-usable macros or compressing existing proofs offline. 
Structural retrieval takes a different approach. Instead of only learning reusable tactics offline, it retrieves structural analogues of the current reasoning situation online—such as proof-state transition motifs, role-aligned fragments, or subgoal evolution patterns—and uses them to guide next-step proof generation.

\paragraph{\textbf{Decomposition, reinforcement learning, and subgoal reasoning}.}

A further line improves proving through decomposition and reinforcement learning. DeepSeek-Prover-V2~\cite{ren2025deepseekproverv2advancingformalmathematical} uses recursive theorem proving, subgoal decomposition, synthetic cold-start data, and reinforcement learning to bridge informal reasoning and formal proof construction in Lean. More recent systems such as Aristotle~\cite{achim2025aristotleimolevelautomatedtheorem} combine Lean proof search with informal reasoning and lemma generation at scale, while Harmonic publicly positions Lean as the backbone of its formal verification pipeline. These approaches show that decomposition is powerful, especially for long-horizon mathematical reasoning. Nevertheless, they mostly address how to generate or search over subgoals, rather than how to retrieve the right reusable structure from past proofs inside a large software-verification corpus. Structural retrieval is therefore complementary: it can serve as the retrieval substrate that supplies decomposition-aware, role-aware, and state-aware analogues from prior proofs.

\paragraph{\textbf{Cross-assistant translation and proof transfer}.}

Some recent efforts explore cross-assistant transfer. lf-lean~\cite{lf-lean} studies Rocq-to-Lean verified translation at scale through automatically generated correctness specifications, and related blog and infrastructure efforts suggest that translation and environment generation may become a practical source of large verified datasets for future training and evaluation. This line is highly relevant to future generalization of our setting across proof assistants. However, the translation objective is still largely instance-level and semantics-preserving. It does not directly model which structural reasoning patterns survive across assistants or across projects. Structural retrieval offers one possible abstraction layer for such transfer, since proof-state transitions, local tactic motifs, and lemma roles may generalize more robustly than surface syntax.

\paragraph{\textbf{Summary}}

In summary, prior work has shown the value of retrieval, decomposition, tactic abstraction, and verifier-guided repair. However, the dominant retrieval granularity remains insufficient levels. We argue that this granularity does not match the true unit of reuse in large-scale verification proofs. By retrieving structural reasoning units rather than only keyword-level retrievals, it aims to provide context that is both finer-grained and more aligned with the current proof state.
\section{Conclusion and Future Works}\label{sec:conclusion}

\sysname\ reframes LLM-assisted automated theorem proving from the perspective of scalability in large-scale system verification. 
While recent approaches achieve strong performance on mathematical benchmarks and small to medium-sized proofs, 
we show that they fundamentally fail to scale to real-world verification projects. 
Our analysis indicates that this limitation does not stem from insufficient reasoning capability of LLMs, 
but from the inability of existing methods to manage deep dependency structures and reuse proof knowledge beyond individual proofs. 
Despite advances in retrieval-augmented and search-based proving, prior techniques remain inherently local, 
operating on static artifacts and single-proof contexts, which limits their effectiveness at system scale. 
We argue that scalable proof automation requires retrieval mechanisms that jointly exploit dependency-level signals and proof-structural regularities across proof attempts. 

As future work, we plan to evaluate our approach on the full seL4 verification corpus to assess scalability in a realistic system setting. 
We also intend to explore learning-based selection strategies, including offline reinforcement learning, 
to replace heuristic retrieval with data-driven policies optimized for long-horizon verification.

\bibliographystyle{ACM-Reference-Format}
\bibliography{refs}

\appendix

\end{document}